# BiEntropy, TriEntropy and Primality

Grenville J. Croll
grenvillecroll@gmail.com

*The order and disorder of binary representations of the natural numbers $< 2^8$ is measured using the BiEntropy function. Significant differences are detected between the primes and the non primes. The BiEntropic prime density is shown to be quadratic with a very small Gaussian distributed error. The work is repeated in binary using a monte carlo simulation for a sample of the natural numbers $< 2^{32}$ and in trinary for all natural numbers $< 3^9$ with similar but cubic results. We find a significant relationship between BiEntropy and TriEntropy such that we can discriminate between the primes and numbers divisible by six. We discuss the theoretical basis of these results and show how they generalise to give a tight bound on the variance of Pi(x) - Li(x) for all x. This bound is much tighter than the bound given by Von Koch in 1901 as an equivalence for proof of the Riemann Hypothesis. Since the primes are Gaussian due to a simple induction on the binary derivative, this implies that the twin primes conjecture is true. We also provide absolutely convergent asymptotes for the numbers of Fermat and Mersenne primes in the appendices.*
*V2.61*

## 1 INTRODUCTION

We developed the BiEntropy function [Croll, 2013] as a means of comparing the relative order and disorder of the digits of binary strings of arbitrary length.

We originally tested the algorithm in the fields of Prime Number Theory, Human Vision, Cryptography, Random Number Generation and Quantitative Finance. As a by-product of our work with prime numbers we derived two very short corollaries which reaffirmed the irrationality [Hardy & Wright, 1979] of the prime constant.

We subsequently used BiEntropy to identify a significant difference between the alternating and non-alternating knots of 9 and 10 crossings [Croll, 2018] in the simple cubic lattice. Our work has been cited in the fields of cryptography, internet information processing, mobile computing and random number generation[Costa et al, 2015][Jin & Zeng, 2015][Jin et al 2016][Kotě et al, 2014][Stakhanova et al, 2016]. Most recently BiEntropy has been re-implemented, tested and made publicly available on GitHub [Helinski, 2018]. It has been prominently cited in a related US Government patent [Gurrieri et al, 2018].

Despite this background of activity on the use and application of BiEntropy in diverse areas, including in particular Prime Number theory, we have failed until now to conduct the simplest of tests to ascertain if there was any relationship between BiEntropy and Primality.

In this paper we empirically investigate the relationship between BiEntropy and Primality in the 8 and 32 bit binary strings. We then develop the TriEntropy function and investigate its relationship with primality within the 9 trit trinary strings. We briefly investigate the relationship between BiEntropy and TriEntropy. We conclude with a discussion of the theoretical basis behind this work and demonstrate how it generalises to all natural numbers.

All of the investigative, experimental and computational work in this paper was performed within the Microsoft Excel spreadsheet environment [Croll, 2005]. This gave us great flexibility in the creative process, high development productivity and notable computational and graphical functionality. These attributes have been already observed [Grossman, 2007 & 2008] and may facilitate the accessibility and furtherance of this work especially within the educational domain [Csernoch & Biró, 2016, 2018].

The layout of this paper reflects the order in which the experimental and theoretical work took place, except that work on the Fermat and Mersenne primes was moved to Appendix 2. We provide online a complete set of spreadsheets used to perform the computations and graphics within this paper. Details regarding access to these spreadsheets is given in the Supplementary Materials section.

## 2. BIENTROPY

The BiEntropy algorithm uses a weighted average of the Shannon Entropies [Shannon, 1948] of a string and all but the last binary derivative [Nathanson, 1971] of the string.

### 2.1 Shannon Entropy

Shannon's Entropy of a binary string $s = s_1, \ldots, s_n$ where $P(s_i=1) = p$ (and $0 \log_2 0$ is defined to be 0) is:

$$H(p) = -p \log_2 p - (1-p) \log_2 (1-p)$$

For perfectly ordered strings which are all 1's or all 0's i.e. $p = 0$ or $p = 1$, $H(p)$ returns 0. Where $p = 0.5$, $H(p)$ returns 1, reflecting maximum variety. However, for a string such as 01010101, where also $p = 0.5$, $H(p)$ also returns 1, ignoring completely the periodic nature of the string.

We can discover the periodicity of a binary string by using the binary derivatives of the string.

### 2.2 Binary Derivatives & Periodicity

The first binary derivative of $s$, $d_1(s)$, is the binary string of length $n - 1$ formed by XORing adjacent pairs of digits. We refer to the $k$th derivative of $s$ $d_k(s)$ as the binary derivative of $d_{k-1}(s)$. There are $n-1$ binary derivatives of $s$. $p(k)$ is the proportion of 1's in $d_k$.

Almost fifty years ago [Nathanson, 1971], following the work of [Goka, 1970] defined the notions of *period* and *eventual period* within arbitrary binary strings and outlined the related properties of binary strings and their derivatives both individually and collectively. Amongst a number of useful results we find that a binary string is periodic with period $2^m$ for some $m \geq 0$ if and only if $d_k = 0$ for some $k \geq 1$.

### 2.3 BiEntropy Definition

BiEntropy, or BiEn for short, is a weighted average of the Shannon entropies of the string and the first $n-2$ binary derivatives of the string. There are numerous ways of weighting the Shannon entropies. In this series of experiments, we weight the Shannon entropies using powers of two:

$$BiEn(s) = \left(1 / (2^{n-1} - 1)\right) \left( \sum_{k=0}^{n-2} ((-p(k) \cdot \log_2 p(k) - (1-p(k)) \cdot \log_2 (1-p(k)))) \cdot 2^k \right)$$

The final derivative $d_{n-1}$ is not used as there is no variation in the contribution to the total entropy in either of its two binary states. The highest weight is assigned to the derivative $d_{n-2}$.

### 2.4 BiEntropy properties

Thus BiEntropy provides a number between 0 and 1 inclusive which indicates the relative order and disorder of the digits of a binary string of length $n > 1$. The shortest perfectly ordered strings are 00 and 11 which have a BiEntropy of 0. The only perfectly disordered strings are 01 and 10 which have a BiEntropy of 1. An ordered (i.e. periodic) string such as 01010101 for example has a low BiEntropy of 0.01. A disordered string such as 10000110 has a high BiEntropy of, e.g., 0.95.

## 3. BIENTROPY & PRIMALITY OF THE NATURAL NUMBERS < 256

We show in Figure 1 below the BiEntropy of the natural numbers < 256. The rows correspond to the most significant digits and the columns the least significant digits of their binary representations. The rows and columns are ordered by the 4 bit BiEntropy of the most and least significant digits respectively. BiEntropy is colour coded with white < 0.15, yellow < 0.25 orange < 0.5 and red < 1.0. Note the symmetry of the diagram about the diagonal. The primes are coloured purple. For example 5 = 00000101 has an 8-bit BiEntropy of 0.23 and would be coloured yellow given the symmetry but is coded purple because it is (a Fermat) prime. The Fermat Prime 17 = 00010001 has a low BiEntropy of 0.05 due to the periodic nature of its digits. It would be coloured white, but is coloured purple because of its primality. 127 = 01111111 has a BiEntropy of 0.92 and would be coloured red, however it is not only prime, but is a Mersenne Prime and is coloured purple. Note that 0 and 1 are simply "not prime".

It is easy to see that most of the primes lie in the red quadrants, with only one prime (a Fermat prime) on the white diagonal. Note that the primality of the natural numbers < 256 has a variance with the natural symmetry of BiEntropy as depicted in Figure 1.

### Table 1 Prime Proportions

| Colour Code | BiEntropy | Count | Prime | Prime Proportion |
|---|---|---|---|---|
| White | < 0.15 | 32 | 1 | 0.0312 |
| Yellow | < 0.25 | 32 | 1 | 0.0312 |
| Orange | < 0.50 | 64 | 15 | 0.2343 |
| Red | < 1.00 | 128 | 37 | 0.2890 |

### Figure 1 – BiEntropy & Primality of the Natural Numbers < 256

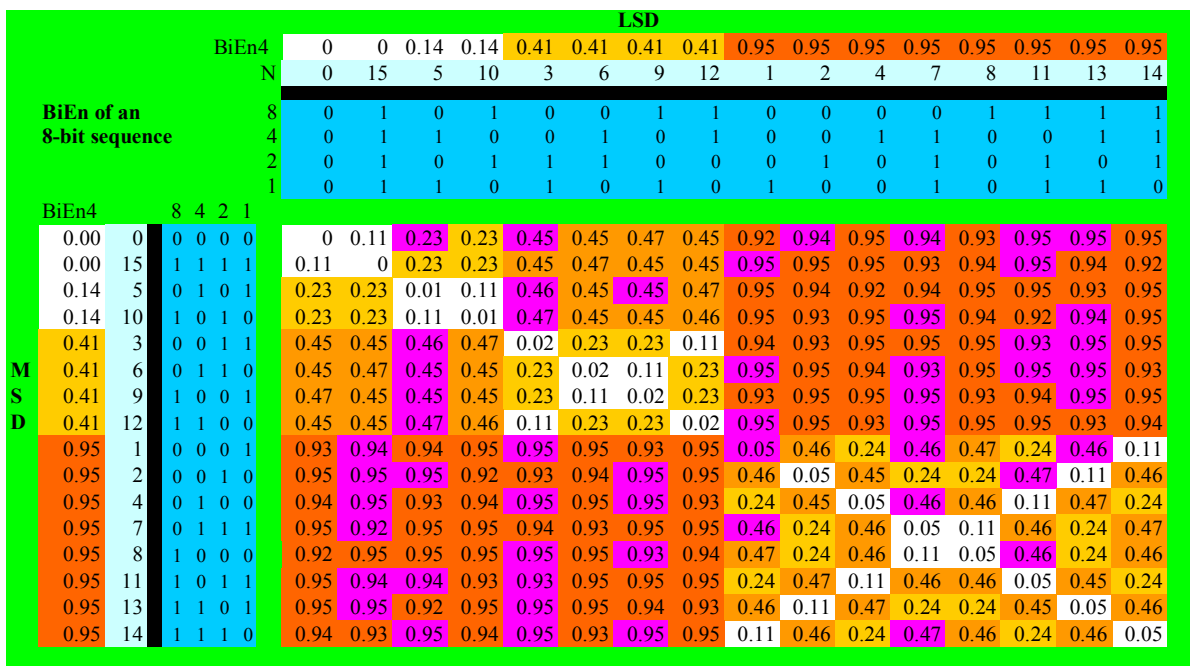

The differences between the four prime proportions of Table 1 above are significant at $p < 0.01$. We have thus discovered a segmentation of the primes based upon BiEntropy, or more generally the binary derivative. Looking for 8 bit primes in the red segment is approximately nine times more productive than looking in the white or yellow segments.

We show in Table 2 below the clear distinction between the BiEntropies of the primes, the non primes and the composite odd numbers at $p < 0.0001$ for the natural numbers < 256. The BiEntropy of the four Mersenne primes < 256 and the 33 twin primes < 256 is similar to the BiEntropy of all the Primes < 256.

Thus the number of the primes and the composite odds < 256 is 129 which includes the even prime.

**Table 2 Mean BiEntropy**

|       | Prime  | Not Prime | Odd    | Mersenne | Twin   |
|-------|--------|-----------|--------|----------|--------|
| Mean  | 0.7897 | 0.5863    | 0.5099 | 0.8134   | 0.7783 |
| S.Dev | 0.2505 | 0.3444    | 0.3497 | 0.2443   | 0.2674 |
| N     | 54     | 202       | 75     | 4        | 33     |

If we sort the natural numbers < 256 by their BiEntropies and group them into 8 segments as in Table 3, differences in prime density between the lowest and highest BiEntropy segment becomes markedly higher.

**Table 3 BiEntropy ordered Prime segments**

| Segment | BiEntropy ≤ | Primes |
|---------|-------------|--------|
| 1       | 0.1141      | 1      |
| 2       | 0.2395      | 1      |
| 3       | 0.4558      | 8      |
| 4       | 0.4734      | 6      |
| 5       | 0.9350      | 6      |
| 6       | 0.9487      | 9      |
| 7       | 0.9506      | 9      |
| 8       | 0.9532      | 14     |

Prime density $\pi(x)$, the number of primes less than or equal to $x$ is approximately $x / \ln(x)$ due to the Prime Number Theorems of Jacques Hadamard and Charles de la Vallée Poussin in 1896. BiEntropy appears to modify the prime density to $O(x^2)$ for very small integers. Using BiEntropy or other prime density functions we can therefore usefully speak of $q(x, y, i)$ which is the number of primes in the $i$th $y$ sized ordered interval < $x$. Thus $q(256, 32, 8)$ is 14 as above. Naturally $\pi(256) = q(256, 256, 1) = 54$.

Finally, we depict the continuous relationship between BiEntropy and Primality graphically in Figure 2, which reveals an almost deterministic relationship. We fit the related natural logarithm and quadratic curves and show the associated errors in Figure 3. We have adjusted the Natural Logarithm curve so that Log(256) matches $\pi(256)$, which $Li(x)$ does in the limit. Note that BiEntropy is a weighted average of the Shannon Entropies of a binary string and the first $n-2$ binary derivatives of the string. No (explicit) trial division has taken place in order to calculate BiEntropy. The number of primes < 256 = 54, and total BiEntropy for the primes < 256 = 42.64.

**Figure 2 BiEntropy Modified prime density**

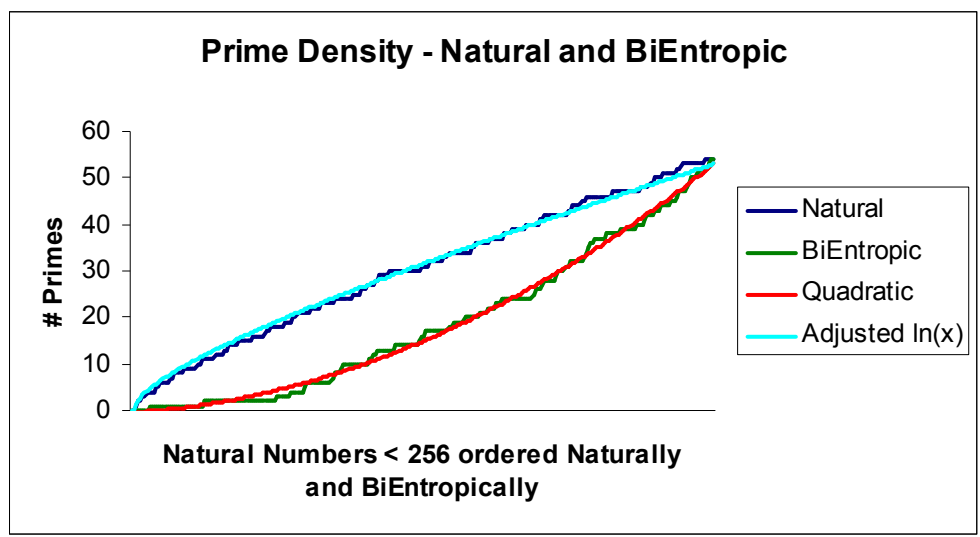

The means of Figure 3 are coincident due to the small multiplicative adjustment we made. The standard deviations of the errors is almost identical, at 0.93 for the natural logarithm and 0.98 for the quadratic. Thus the actual error in the BiEntropic prime density for integers $x < 256$ is $< \sqrt{x} \log(x)$ and is evidently Gaussian. As we will see, the error converges as $x \to \infty$.

### Figure 3 Variation in prime density

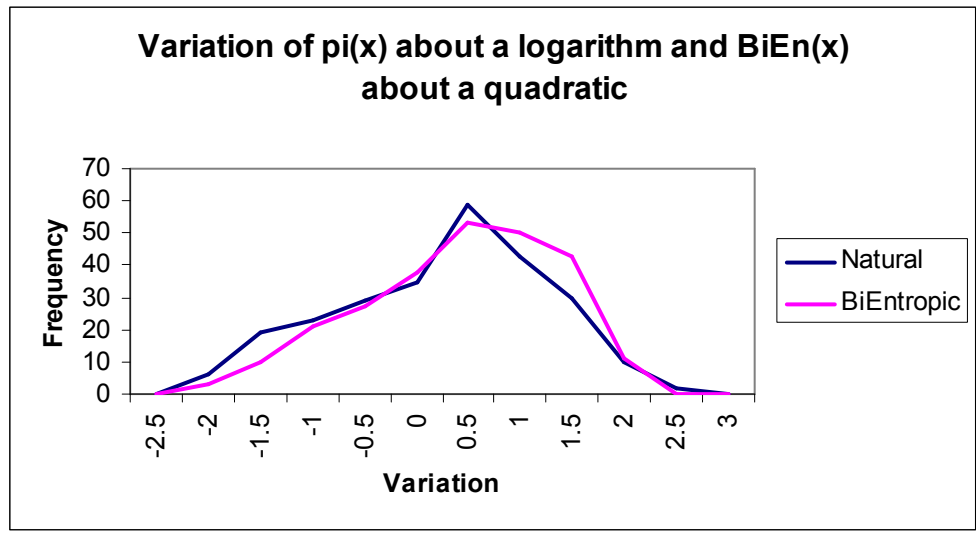

## 4. BIENTROPY & PRIMALITY OF THE NATURAL NUMBERS < $2^{32}$

### 4.1 Primes and Binary Derivatives

Whereas $\pi(x) \sim x / \ln(x)$, the number of binary derivatives used in the calculation of BiEntropy of a string of length $m$ (where $m = \log_2(x)$) increases only as $(m^2 - m) / 2$. We show in Table 4 the relationship between $\pi(x)$, the number of primes and the binary derivatives $d$ for various $x$.

### Table 4 - $\pi(x)$ and the number of binary derivatives for various $x$.

| $x$ | Bits($m$) | $\pi(x)$ | Derivatives($d$) | d/$\pi(x)$ % |
|---|---|---|---|---|
| 256 | 8 | 54 | 28 | 51.85% |
| 65536 | 16 | 6542 | 120 | 1.83% |
| 4294967296 | 32 | 203280221 | 496 | 0.00% |

Thus $d / \pi(x)$ tends to zero very rapidly, potentially rendering BiEntropy less sensitive to primality at longer string lengths.

### 4.2 Higher Powers of Shannon Entropy

There is some research originating in algorithmic information theory [Chaitin & Schwarz, 1978] that suggests that primality is related to disorder, which is of course what BiEntropy is designed to measure. This other work does not address the use of the binary derivative in this process. Note that there is only one prime on the diagonal of Figure 1, which is the region of maximum order.

We can change BiEntropy to sharpen its sensitivity to the detection of any departures from perfect disorder in the binary derivatives. This is trivially easy to do, especially within the spreadsheet environment, as we can simply and easily raise the Shannon Entropy of each binary derivative to a power higher than 1.

We show in Figure 4 the effect of raising the powers of Shannon Entropy from 1 to 10 based upon $p$, the variety. In the region of maximum variety in the middle of the chart where the variety, $p = 0.5$, the Shannon Entropy is highest. When using higher powers of Shannon Entropy we can more powerfully discriminate departures from maximal disorder.

**Figure 4 Raising Shannon Entropy to a Higher Power**

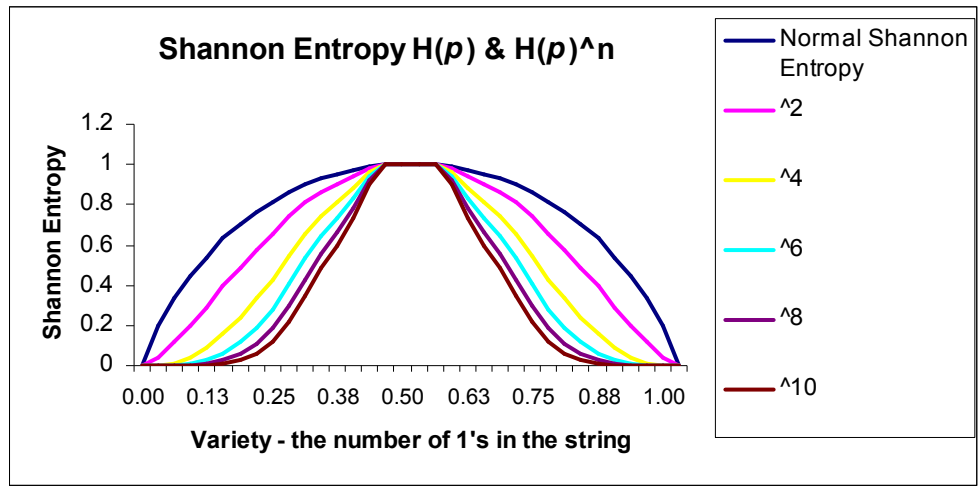

### 4.3 Investigating BiEntropy and Primality for $x < 2^{32}$

We used a spreadsheet based monte carlo calculation to investigate a sample of natural numbers $< 2^{32}$. Using a simple Excel data table, for each of 10,000 iterations, we generated a random 32 bit integer and then calculated its quadratic BiEntropy using the tenth power of the Shannon Entropy of each derivative (P10 BiEntropy). We used a spreadsheet based exhaustive trial division calculation to determine the primality of each random 32 bit integer. We then sorted the sampled natural numbers and their BiEntropies into BiEntropic order and compared the prime density of this ordered interval with the natural prime density of the sample. We show the relationship between the sample's natural and BiEntropic prime density in Figure 5 and the difference between the two densities in Figure 6.

### 4.4 Testing the BiEntropy and Primality Monte Carlo

We decided to carefully investigate the small difference between the Natural and BiEntropic prime density generated by the monte carlo simulation as depicted in Figure 5 below. The simulation consisted of 10,000 samples of an integer $x$ in the range $0 < x < 2^{32}$ produced by the Excel RAND function, which we have previously scrutinised [Croll, 2013]. Since random numbers would be generated uniformly (i.e. linearly) in the given range, we were able to calculate, using the Prime Number Theorem, how many primes were likely to be produced during the generation of 10,000 random integers in the given range. We were then able to calculate a theoretical prime density for the monte carlo simulation to compare against the actual prime density of the monte carlo simulation.

We show the actual difference or Delta between the Natural and BiEntropic Prime Densities with the theoretical expected difference or Delta in Figure 6. The theoretical Delta, shown in orange, accounts for only part of the difference. The difference between the BiEntropic prime density and the natural prime density is not accounted for by the difference between the linear production rate of prime numbers in the monte carlo simulation and the natural prime density. The difference is much larger. Squaring the expected difference and dividing by two (Delta$^2$ / 2) matches the actual results of the monte carlo simulation much more closely. By examination the error between theoretical and actual in Figure 6 is broadly normal (mean 1.22 & St. Dev. 6.17) and omitted for brevity. It appears to be case that BiEntropic Prime Density is also quadratic for integers of $O(2^{32})$.

### Figure 5 P10 BiEntropy and prime density

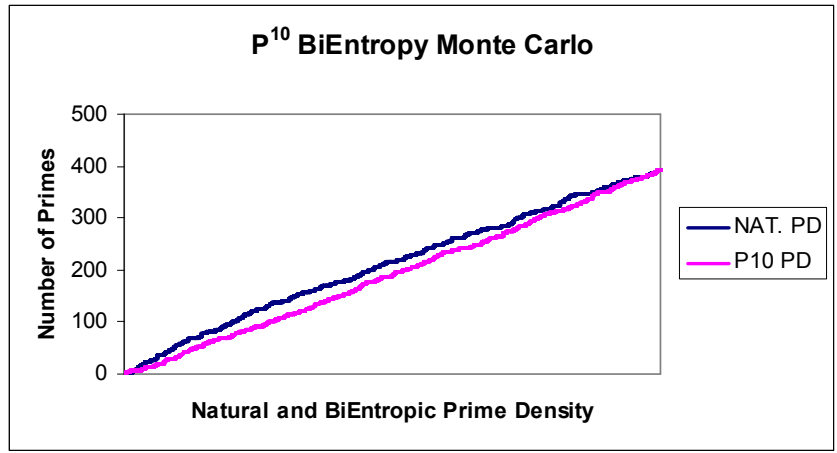

The number of primes actually produced in the monte carlo simulation we have reported is 391 compared with 473 expected prime numbers. A variation was to be expected

### Figure 6 P10 BiEntropy and Prime Density Delta

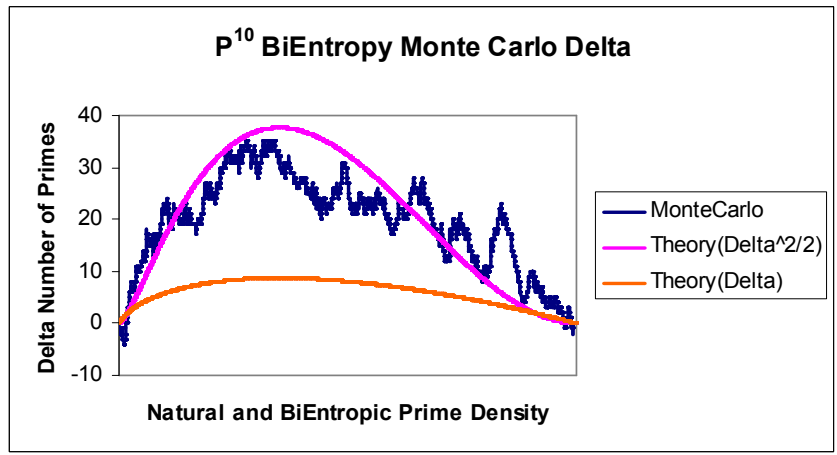

## 5 TRIENTROPY

We have noticed in previous work that the BiEntropy function is not sensitive to periodicities of 3 (see the entries for 18, 27, 36 & 54 in Appendix 1). For example the 18 bit quadratic BiEntropy of 001001001001001001 is 0.9484 indicating disorder, however the string is clearly periodic. We had thought that developing a trinary equivalent to BiEntropy might fix this problem, but were not previously motivated to do so. Given the association between BiEntropy and primality outlined in the previous sections, and the fact that all primes ≥ 5 are of the form $6k \pm 1$, there became a clear motive to investigate.

### 5.1 Pairwise Addition & Differences Modulo 3

The acid test for TriEntropy was that it picked up periodicities of 3 within a Trinary string. We devised a simple two way pairwise trinary addition table which we illustrate in Table 5.

#### Table 5 Pairwise Trinary Addition Table

|   | 0 | 1 | 2 |
|---|---|---|---|
| **0** | 0 | 1 | 2 |
| **1** | 1 | 0 | 1 |
| **2** | 2 | 1 | 0 |

We transformed our 8 bit binary BiEntropy calculator spreadsheet into a 9 trit TriEntropy calculator spreadsheet using the pairwise trinary addition table of Table 5 above. This took just a few minutes. Unfortunately it did not work. We then discovered that in a 3 trit trinary string *ABC* we needed to compute the three way Pairwise Trinary Differences (PTD) between the three pairs *AB, BC, AC,* modulo 3. Thus

$$PTD = MOD(ABS(A-B) + ABS(B-C) + ABS(A-C), 3)$$

which we show in Table 6. It took a few more minutes in a spreadsheet to show that this did indeed work. The respective TriEntropies of the three trit strings did not look promising, however we persisted with our analysis. Note that the PTD function is invariant under pairwise permutation, none of *A, B* or *C* having priority.

**Table 6 Pairwise Trinary Difference (PTD) Table**

| A | B | C | PTD | TriEntropy |
|---|---|---|-----|------------|
| 0 | 0 | 0 | 0 | 0.168 |
| 0 | 0 | 1 | 2 | 0.395 |
| 0 | 0 | 2 | 1 | 0.395 |
| 0 | 1 | 0 | 2 | 0.395 |
| 0 | 1 | 1 | 2 | 0.395 |
| 0 | 1 | 2 | 1 | 0.395 |
| 0 | 2 | 0 | 1 | 0.395 |
| 0 | 2 | 1 | 1 | 0.395 |
| 0 | 2 | 2 | 1 | 0.395 |
| 1 | 0 | 0 | 2 | 0.395 |
| 1 | 0 | 1 | 2 | 0.395 |
| 1 | 0 | 2 | 1 | 0.395 |
| 1 | 1 | 0 | 2 | 0.395 |
| 1 | 1 | 1 | 0 | 0.168 |
| 1 | 1 | 2 | 2 | 0.395 |
| 1 | 2 | 0 | 1 | 0.395 |
| 1 | 2 | 1 | 2 | 0.395 |
| 1 | 2 | 2 | 2 | 0.395 |
| 2 | 0 | 0 | 1 | 0.395 |
| 2 | 0 | 1 | 1 | 0.395 |
| 2 | 0 | 2 | 1 | 0.395 |
| 2 | 1 | 0 | 1 | 0.395 |
| 2 | 1 | 1 | 2 | 0.395 |
| 2 | 1 | 2 | 2 | 0.395 |
| 2 | 2 | 0 | 1 | 0.395 |
| 2 | 2 | 1 | 2 | 0.395 |
| 2 | 2 | 2 | 0 | 0.168 |

**5.2 Computing TriEntropy**

In order to compute the Shannon Entropy of a Trinary string we need the $p_i$ of all the possible symbols. For the derivatives, as in Table 6 above, the $p_i$ for 0, 1, 2 are 0.111 (3/27), 0.444 (12/27) and 0.444 (12/27) respectively. Importantly, since TriEntropy would necessarily calculate the Shannon Entropy of the original string note that the $p_i$ of the 0,1,2 of the input string are 0.333, 0.333, 0.333 as they are equiprobable. Furthermore, note that only $(n-1)/2 - 1$ derivatives are possible (where $n$ is odd) as three input trits are required to compute each output trit of the derivatives. Finally note that in BiEntropy, once a periodicity is detected, the further derivatives automatically fall to zero. This is not the case for TriEntropy and so derivatives that fall to 0 have to have their further non-use programmed in specifically. Note $n$ is odd.

$$TriEn(s) = \left(1 / \left(\sum_{k=0}^{(n-1)/2} 3^k\right)\right) \left(\sum_{k=0}^{(n-1)/2} (-p(k) \cdot \log_2 p(k) - (1 - p(k)) \cdot \log_2 (1 - p(k))) \cdot 3^k\right)$$

We show in Table 7 the layout of a simple Excel spreadsheet to compute the polynomial (ie cubic) TriEn of a 9 trit string. We used Table 6 above to compute each trit of the derivatives.

## Table 7 Computing 9-Trit TriEntropy

| Trinary Expansion of N | Len(N) | N0 | N1 | N2 | p | (1-p) | -p.log(p) | -(1-p).log(1-p) | TriEn | k | 3^k | TriEn*3^k |
|---|---|---|---|---|---|---|---|---|---|---|---|---|
| 1 1 1 2 0 1 1 0 1 | 9 | 2 | 6 | 1 | 0.33 | 0.67 | 0.53 | 0.39 | 0.92 | 0 | 1 | 0.92 |
| 0 2 1 1 2 2 2 | 7 | 1 | 2 | 4 | 0.40 | 0.60 | 0.53 | 0.44 | 0.97 | 1 | 3 | 2.91 |
| 1 2 2 2 0 | 5 | 1 | 1 | 3 | 0.38 | 0.62 | 0.53 | 0.43 | 0.96 | 2 | 9 | 8.61 |
| 2 0 1 | 3 | 1 | 1 | 1 | 0.33 | 0.67 | 0.53 | 0.39 | 0.92 | 3 | 27 | 24.79 |
|  |  |  |  |  |  |  |  |  | 3.76 | 6 | 40 | 37.23 |
|  |  |  |  |  |  |  |  |  | TriEn |  |  | 0.93 |

We exhaustively calculated TriEn for all $x < 3^9$ and show the resulting Natural and TriEntropic prime densities in Figure 7. The equivalent BiEntropy diagram for all $x < 2^{16}$ is almost identical and was earlier omitted for brevity. We show the difference or delta between TriEntropic Prime density and Natural Prime Density in Figure 8.

### Figure 7 TriEntropy and prime density

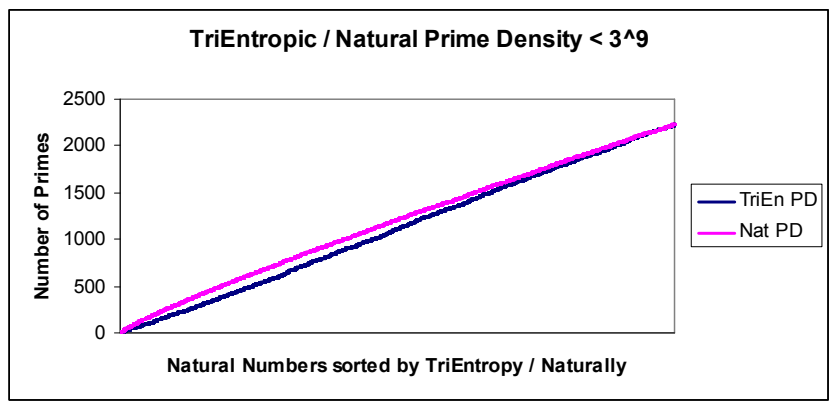

### Figure 8 TriEntropy and prime Density Delta

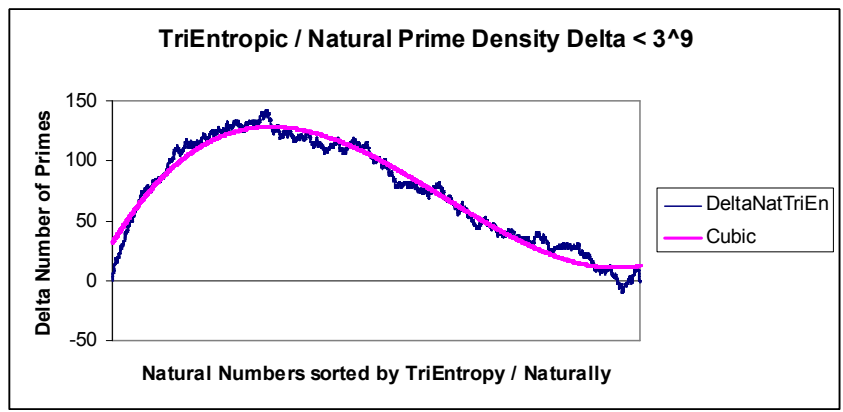

Thus the difference between the Natural and TriEntropic prime densities for $x < 3^9$ is approximately cubic. The error of the difference is approximately Gaussian, which we depict in Figure 9. Mean error is 0.00 with a standard deviation of 7.34.

**Figure 9 TriEntropy Delta / Cubic Error**

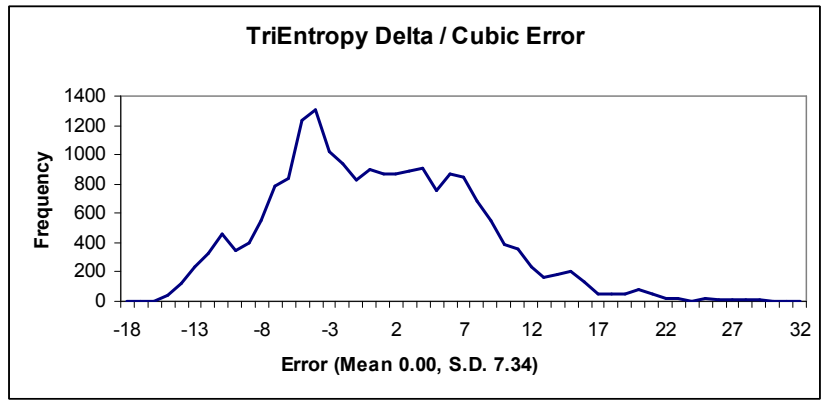

## 6 INTERACTION BETWEEN BIENTROPY AND TRIENTROPY

We investigated the interaction between BiEntropy and TriEntropy in the Natural numbers < 256. We did this by allocating two segment numbers between 0 and 15 inclusive to each natural number depending upon the BiEntropy and TriEntropy. The 16 natural numbers with the lowest BiEntropy were allocated to BiEntropy segment 0 and the 16 natural numbers with the highest BiEntropy were allocated to BiEntropy Segment 15 etc and similarly for TriEntropy. We show in Figure 10 below a diagram of the frequency of occurrence of the primes in blue and numbers divisible by six in red arranged by BiEntropic segment number on the $x$ axis and by TriEntropic segment number on the $y$ axis. The primes are coded as positive numbers and the numbers divisible by six are coded as negative numbers. There was one collision in segment 8-9 corresponding to the numbers 42 and 103 which is coded yellow.

Although the data volume is small, we expect from our earlier experiments that increasing BiEntropy and increasing TriEntropy will disclose more primes and fewer composites. This is what appears to be the case. Ignoring the bottom left to top right diagonal, primes are relatively absent from the top left triangle (11 / 120 versus 40 / 120, $p < 0.0001$) and numbers divisible by six are relatively absent from the bottom right triangle (11 / 120 versus 30/120, $p < 0.002$), which corresponds to prior expectation.

**Figure 10 BiEntropy & TriEntropy interaction < 256**

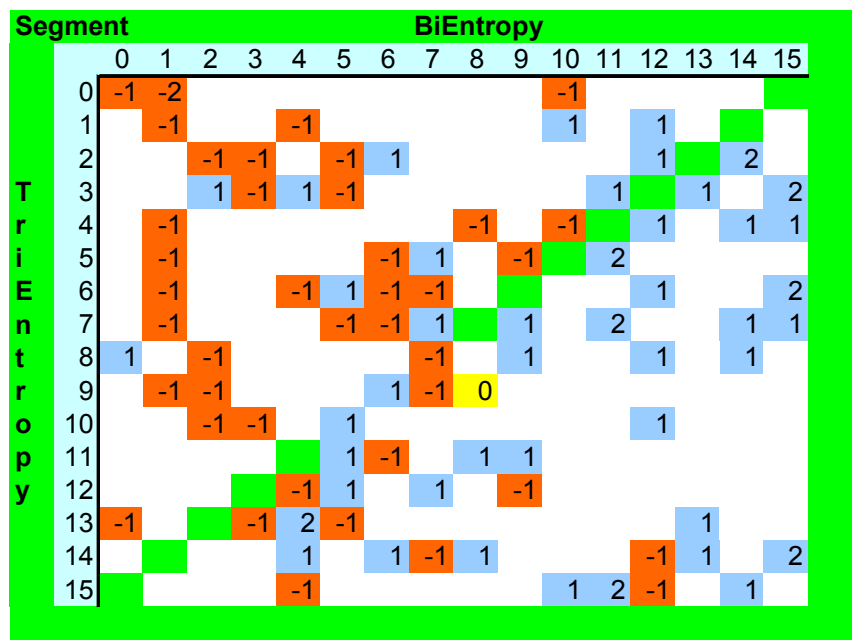

There was only one segment collision, whereas 8 might have been expected (54 * 42 / 256) if the distribution of primes and numbers divisible by six was uniform across all the BiEntropic and TriEntropic segments. Note that the 202 non-primes are uniformly distributed across Figure 10, which information is not shown for brevity, but is available in the Supplementary Materials.

## 7 THEORETICAL BASIS

### 7.1 Introduction

We now show why the notion of periodicity is so critically important in the determination of primality.

### 7.2 Periodic & Non-Periodic Numbers

Consider the concatenated binary string *ab* where the length of *a* and *b* is *n* and $n \geq 1$, so the length of *ab* is 2*n*. If *a = b*, for some *n* then *ab* is periodic. The periodic numbers appear along the diagonal emanating from the origin of Figure 1 (where *n* = 4) and are mostly coloured white.

### 7.3 Periodic Binary Primes

Where *a = b =* 1, some of these are the Fermat numbers, of which only five are known to be prime [Boklan & Conway, 2016]. A Fermat number, 17, appears on the diagonal of Figure 1 and is coloured purple because it is prime. We discuss the Fermat numbers in more detail in Appendix 2.

### 7.4 Periodic Binary Composites

The rest of the numbers, *k*, on the diagonal emanating from the origin of Figure 1 and its equivalents for all *n* are of the form:

$$k = (2^n \cdot a) + b$$

since  *a = b*

then  $k = (2^n + 1) \cdot a$

since  *a > 1*

therefore  *k* is composite.

The first periodic binary composite > 0 is 1010 which is 10 (ten). Thus the Mersenne numbers (numbers of the form $2^n - 1$) of even length (where *a = b* ) cannot be prime. The odd length Mersenne numbers eg 0111, seven, may be prime but are not periodic because $a \neq b$. We list the periodic binary composites < 256 in Appendix 1 and discuss the Mersenne numbers in more detail in Appendix 2.

### 7.5 N-Periodic Binary Composites

Numbers of the form 00111100 and 10010110 etc where *a* is the 2's complement of *b* ie

$$b = 2^n - a - 1$$

are also composite. These numbers appear in white in the short cross diagonals of Figure 1. These numbers are also of the form:

$$k = (2^n \cdot a) + b$$

Substituting $\quad k = (2^n . a) + 2^n - a - 1$

Therefore $\quad k = a.(2^n - 1) + 2^n - 1$

$\quad k = (a + 1).(2^n - 1)$

if $\quad a \geq 1$ then $k$ is composite

else $\quad a = 0$ and $b$ is a periodic binary composite (eg 1111…) of length $n / 2$

### 7.6 Periodic M-ary Primes

All primes $k > 2$ are of the form $ab$ where $a$ and $b$ are of length $n$ and $a = b = 1$ in one base $m = k - 1$.

since $\quad a = b = 1$

then $\quad k = ((k-1)^1 . 1) + 1$

Except that the Fermat primes are also periodic in base 2.

That is $\quad k = (2^{(n-1)}.a) + b$ where $a = b = 1$.

### 7.7 Periodic M-ary Composites

In general, the numbers $k$, on the diagonals of diagrams equivalent to Figure 1 in any base $m$ are of the form:
$\quad k = (m^n . a) + b$

since $\quad a = b$

then $\quad k = (m^n + 1) . a$

since $\quad a > 1,\ k$ is composite.

### 7.8 Non periodic numbers

Numbers where $a \neq b\ (n \geq 1)$ in any base are either prime or non-prime.

## 8 DISCUSSION

Thus the principal reason why BiEntropy & TriEntropy have any relationship with primality is the simple fact that, save the Fermat numbers (and e.g. $23 = 11_{22}$), periodic and $n$-periodic numbers cannot be prime in any base. Hence the main diagonal of Figure 1 (for all $x$, and in all bases) is almost devoid of primes and there are no primes on the cross diagonals. Ignoring the Fermat primes, $32 / 256 = 12.5\%$ of natural numbers $< 256$ cannot be prime due to the periodicity or $n$-periodicity within seven of their last eight binary derivatives.

If a binary string is periodic, one, and then all the further derivatives, fall to zero [Nathanson, 1971]. BiEntropy picks this up as the Shannon Entropy is zero. Symmetrically, if a derivative is all 1's, it will also have a zero Shannon entropy and will (unless it is the last used derivative) become all zeroes in the next derivative. The earlier that periodicity is observed (i.e. for shorter periods) the lower the weighted total becomes as all the higher weights are zero. Non-periodic strings are otherwise ranked accordingly, with those strings with the most derivatives at or close to $p = 0.5$ gaining the highest BiEntropy. BiEntropy is the Hamming distance for primality. Except in certain circumstances (eg $s = 00000001$), the bits of a binary derivative are undecideable. Determination of the bits of a binary derivative is a simple variation of the halting problem – if the last binary derivative is one the routine halts else it does not halt.

Whereas a string is periodic if and only if one of its derivatives is all zeroes, the reverse does not apply, hence primality is stochastic. [Davies et al, 1995] prove that if the bits of a string occur with a probability of 0.5, the bits of the derivatives also occur with a probability of 0.5, and the binary derivatives are independent. Hence the error between a quadratic and the BiEntropic prime density, which is a quadratic function of the probability of occurrence of each bit of the derivative, is Gaussian due to the central limit theorem. Note that the number of binary derivatives for any $x$ is finite.

BiEntropic prime density is quadratic because BiEntropy is quadratic. For example in the 8 bit version of BiEntropy, the probability of an arbitrary string not being prime because the string is all 1's or it or a binary derivative is zero is:

$$P(s \text{ is not prime}) = 1/256 + 1/256 + 2/256 + 4/256 + 8/256 + 16/256 + 32/256 + 64/256$$

$$= 128/256 = 0.5$$

I.e., there is only one 8 bit string ($s = d_0$) that is all zeroes, whereas the last used derivative $d_6$ of length 2 is all zeroes on 64 occasions and $d_5$ of length 3 is all zeroes on 32 occasions. BiEntropy measures *exactly* the probability that a binary string *cannot* be prime or *may be* prime with a precision given by the number of bits in $d_0$. This probability is independent of the Prime Number Theorem. TriEntropy is cubic for similar reasons.

By a simple induction, every binary number is the binary derivative of a number one bit longer, and therefore its bits occur with a probability of 0.5. Its bits are proven [Davies et al, 1995] independent of its earlier derivatives. Hence, the primes are Gaussian, as the probability of occurrence of each of their bits is undifferentiated from every other binary number and every other binary derivative.

The relationship between BiEntropy and Primality generalises for all $x$ for the simple reason that all $x \geq 256$ (say) eventually end up as an 8 bit (say) string by virtue of successive binary differentiation. Determination of many of the mathematical and statistical properties of all $x$ can be obtained inductively by observation of properties in the last $m$ binary derivatives, which is easy to do when $m$ is small.

Thus, there exists a set of constants $a_k$, $b_k$ & $c_k$ such that

$$a_k.x_k^2 + b_k.x_k + c_k = Li(x_k) \text{ where } x_k = m^2, m \text{ is integer and } c_k = 0$$

And another (similar) set of constants $u_k$, $v_k$ & $w_k$, such that

$$u_k.x_k^2 + v_k.x_k + w_k = \pi(x_k) \text{ where } x_k = m^2, m \text{ is integer and } w_k = 0$$

For each $a_k$, $b_k$, $c_k$ and $u_k$, $v_k$ & $w_k$, there exists a set of $(m^2 - m)/2$ binary derivatives from which the distribution of primes is derived with known probabilities and calculable or estimable variance. The variance in natural prime density is constrained by the variance in the BiEntropic prime densities for all $x_k < x$ because the same data – the natural numbers - is Gaussian distributed about two differing central measures – a quadratic and a logarithmic integrand.

Since $$\lim_{x \to \infty} \pi(x)/Li(x) \sim 1$$

Therefore, in the limit, the BiEntropic / Quadratic and Logarithmic Integrand / Natural error distributions are coincident with near identical error distributions, which we illustrated empirically in Figure 3. Furthermore, as $x \to \infty$, and since the number of bits in the binary

derivatives = $(m^2 - m) / 2$, where $m = \log_2(x)$, the variance in the error between the BiEntropic and Quadratic prime densities is $O(\log(x) / x)$ due to the central limit theorem. Hence, the error between the Logarithmic Integrand and the natural prime density rapidly tends to 0.

ie $\qquad \lim_{x \to \infty} \text{Var}(\pi(x) - Li(x)) \to 0$

Which is clearly distinctive from the [von Koch, 1901] bound for the proof of the Riemann Hypothesis.

$$\pi(x) - Li(x) = O(\sqrt{x} \log(x))$$

A similar set of cubic constants apply for TriEntropy and the arithmetic addition of BiEntropy and TriEntropy, which we shall denote TriBiEntropy. We illustrate the cubics of TriBiEntropy intersecting $\pi(x)$ for various $x$ in Figure 11.

**Figure 11 BiEntropy + TriEntropy and $\pi(x)$ for various $x < 16k$**

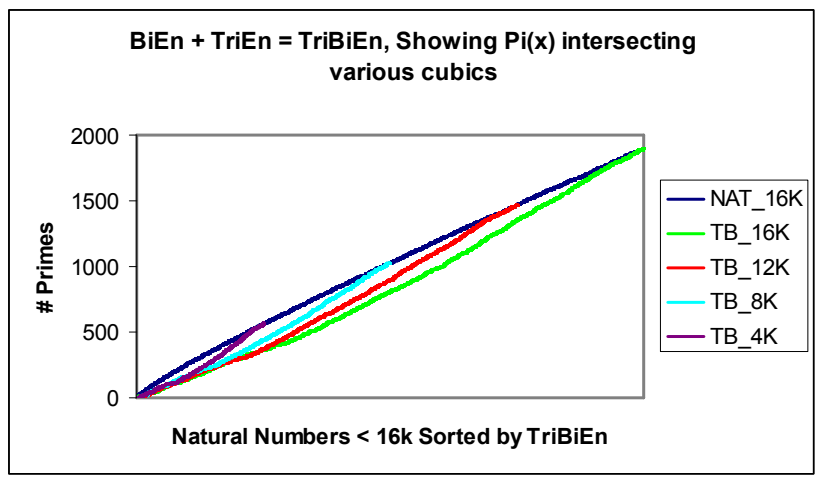

## 9. CONCLUSIONS

We have shown a clear empirical link between BiEntropy and primality for the natural numbers $< 2^8$. We have repeated this analysis statistically for the natural numbers $< 2^{32}$ and found similar results, including the prime density remaining $O(x^2)$. We developed a related TriEntropy function and show that TriEntropy changes prime density to $O(x^3)$ for the natural numbers $< 3^9$. In addition, TriEntropy has addressed a natural weakness in the detection of periods of length three or multiples thereof in the BiEntropy function.

Since BiEntropy and TriEntropy are simply measures of the order and disorder (i.e. the periodicity) of a string, the implication is that prime numbers expressed in binary or trinary have more disordered representations. The reverse implication is that composites have more ordered representations. This result has been suggested in earlier work in algorithmic information theory.

We have shown how to increase the sensitivity of BiEntropy by increasing the exponent of the Shannon Entropy within the BiEntropy calculation. We have demonstrated a significant link between BiEntropy and TriEntropy in the natural numbers $< 2^8$ and the practicality of combining BiEntropy and TriEntropy via arithmetic addition. We have given a brief outline of the theoretical basis behind this initial experimental work and show how it generalises for all the natural numbers.

We have shown how the variance of the error between $\pi(x)$ and $Li(x)$ tends to zero due to the Gaussian constraints on the variance of $\pi(x)$ imposed by the binary derivative. These are much tighter constraints than the bound proven by Von Koch in 1901 as being equivalent to proof of the Riemann Hypothesis. We have provided in Appendix 2, two easily derived absolutely convergent asymptotes for the numbers of Fermat and Mersenne primes.

Finally, since the distribution of primes is Gaussian due to the binary derivative, this implies that the twin primes conjecture is true.

**FURTHER WORK**

Note from Figure 1 and earlier work, that BiEntropy, TriEntropy etc though reals are quantized and not continuous, there being a finite number of states. This may be of relevance in attempts to relate BiEntropy & TriEntropy to the physical domain [Croll, 2014]. There are myriad opportunities to relate primality to other domains [Guariglia, 2019], particularly bearing in mind the now established connections between binary, the binary derivative, primality, and their *m*-ary generalisations.

**SUPPLEMENTARY MATERIALS**

The full set of spreadsheets used to perform the monte carlo simulations, computations, tables and graphics of this paper is available in the Figshare Data Depository: Croll, Grenville (2020), BiEntropy_TriEntropy_and_Primality.zip. figshare. Dataset. https://doi.org/10.6084/m9.figshare.11743749

**ACKNOWLEDGEMENTS**

I thank my wife for her support and for part financing this work and its publication January 2019 – March 2020. I thank my colleagues at ANPA, PANPA and EuSpRIG for their advice, contributions, companionship and support. I thank my family and friends for tolerating with good humour my mathematical monologues. There are no conflicts of interest to declare. In loving memory of my parents.

## Appendix 1 – Periodic Binary Composites < 256

Periodic binary numbers of the form *ab* where *a = b ≠ 1* cannot be prime.

| Bits | Binary    | Decimal | BiEntropy |
|------|-----------|---------|-----------|
| 4    | 1010      | 10      | 0.14      |
| 4    | 1111      | 15      | 0.00      |
| 6    | 010010    | 18      | 0.44      |
| 6    | 011011    | 27      | 0.95      |
| 8    | 00100010  | 34      | 0.05      |
| 6    | 100100    | 36      | 0.95      |
| 6    | 101101    | 45      | 0.44      |
| 8    | 00110011  | 51      | 0.02      |
| 6    | 110110    | 54      | 0.95      |
| 6    | 111111    | 63      | 0.00      |
| 8    | 01000100  | 68      | 0.05      |
| 8    | 01010101  | 85      | 0.01      |
| 8    | 01100110  | 102     | 0.02      |
| 8    | 01110111  | 119     | 0.05      |
| 8    | 10001000  | 136     | 0.05      |
| 8    | 10011001  | 153     | 0.02      |
| 8    | 10101010  | 170     | 0.01      |
| 8    | 10111011  | 187     | 0.05      |
| 8    | 11001100  | 204     | 0.02      |
| 8    | 11011101  | 221     | 0.05      |
| 8    | 11101110  | 238     | 0.05      |
| 8    | 11111111  | 255     | 0.00      |

# APPENDIX 2 THE FERMAT & MERSENNE PRIMES

A Fermat number is of the form $2^{2^n} + 1$. Fermat numbers are also periodic binary numbers of the form $ab$ where $a = b = 1$. We show in Table 8 the five known Fermat primes.

### Table 8 – The Fermat Primes

| Fermat Number | Decimal | Binary | n |
|---|---|---|---|
| $F_0$ | 3 | 11 | 1 |
| $F_1$ | 5 | 0101 | 2 |
| $F_2$ | 17 | 00010001 | 4 |
| $F_3$ | 257 | 0000000100000001 | 8 |
| $F_4$ | 65537 | 00000000000000010000000000000001 | 16 |

The prime number theorems of Jacques Hadamard and Charles de la Vallée Poussin in 1896 infer that for a binary string $x$ of length $2n$ bits, the probability, $p$, of it being prime is $\sim 1 / \log(2^{2n})$. There are $2^{2n}$ binary strings of length $2n$, and one and only one of which can be a periodic binary prime of period length $n$ where $a = b = 1$.

Thus the probability that a binary string of length $2n$ is a prime of period $n$ is

$$p(F_n = \text{prime}) = 1/ \log (2^{2n}+1)$$

Note that the asymptote of the Prime Number Theorem accounts for even numbers and periodic numbers which cannot be prime. Since periodicity and hence potential primality is an independent function of each string of length $2n$, the total number of possible Fermat primes

$$F_\infty = \sum_{n=1}^{n=\infty} (1 / \log (2^{2n}+1))$$

Which by examination in Table 9 & Figure 12 converges to about 5 for $2n < 2^{1024}$ (the spreadsheet floating point limit) and is absolutely convergent by D'Alembert's criterion.

### Figure 12 – Actual and Expected Fermat Primes

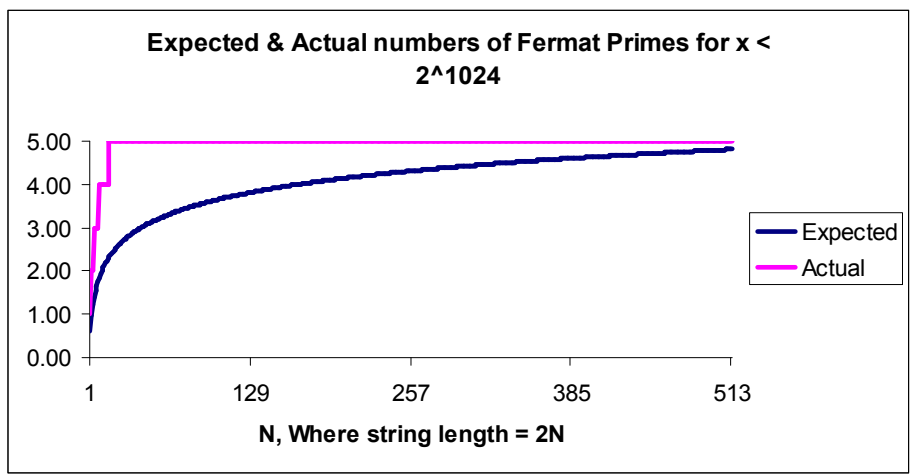

Continuation of the series by a log approximation reveals a 6th Fermat prime by the 3226th term and a 7th Fermat prime by the 13,651st term. But it has been shown that for numbers of the form $2^k + 1$ which are prime, $k$ is a power of two. As [Boklan and Conway, 2016] discuss, further Fermat primes are highly unlikely, despite the continuation of the log approximation

**Table 9 – Actual and Expected Fermat Primes**

| N | $2^{2n}$ | $2^n+1$ | ln(2n) | $\sum p(F(n))$ | F | D'Alembert's criterion |
|---|---|---|---|---|---|---|
| 1 | 4 | 3 | 0.6213 | 0.6213 | 1 | |
| 2 | 16 | 5 | 0.3530 | 0.9743 | 2 | 0.5681 |
| 3 | 64 | 9 | 0.2396 | 1.2138 | 2 | 0.6787 |
| 4 | 256 | 17 | 0.1802 | 1.3941 | 3 | 0.7523 |
| 5 | 1024 | 33 | 0.1442 | 1.5383 | 3 | 0.8004 |
| 6 | 4096 | 65 | 0.1202 | 1.6585 | 3 | 0.8334 |
| 7 | 16384 | 129 | 0.1030 | 1.7616 | 3 | 0.8572 |
| 8 | 65536 | 257 | 0.0902 | 1.8517 | 4 | 0.8750 |
| 9 | 262144 | 513 | 0.0801 | 1.9319 | 4 | 0.8889 |
| 10 | 1048576 | 1025 | 0.0721 | 2.0040 | 4 | 0.9000 |
| 11 | 4194304 | 2049 | 0.0656 | 2.0696 | 4 | 0.9091 |
| 12 | 16777216 | 4097 | 0.0601 | 2.1297 | 4 | 0.9167 |
| 13 | 67108864 | 8193 | 0.0555 | 2.1852 | 4 | 0.9231 |
| 14 | 268435456 | 16385 | 0.0515 | 2.2367 | 4 | 0.9286 |
| 15 | 1073741824 | 32769 | 0.0481 | 2.2848 | 4 | 0.9333 |
| 16 | 4294967296 | 65537 | 0.0451 | 2.3299 | 5 | 0.9375 |
| 17 | 17179869184 | 131073 | 0.0424 | 2.3723 | 5 | 0.9412 |
| 18 | 68719476736 | 262145 | 0.0401 | 2.4124 | 5 | 0.9444 |
| 19 | 2.74878E+11 | 524289 | 0.0380 | 2.4504 | 5 | 0.9474 |
| 20 | 1.09951E+12 | 1048577 | 0.0361 | 2.4864 | 5 | 0.9500 |

A similar argument prevails for the Mersenne primes except that there are two terms accounting for numbers of the form 01111111 and 11111111. The prime number theorem does not "know" that even length periodic numbers cannot be prime. The series is again absolutely convergent, but converges much more slowly. The total number of Mersenne primes is therefore:

$$M_\infty = \sum_{n=1}^{n=\infty} 1/\log(2^{(2n-1)} - 1) + 1/\log(2^{(2n-2)} - 1)$$

**Figure 13 – Actual and Expected Mersenne Primes**

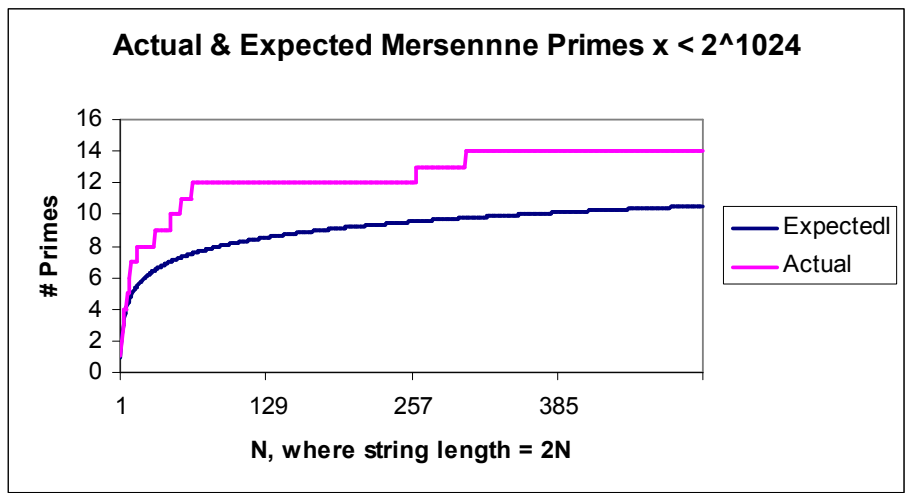

There are presently 51 known Mersenne primes, $2^{82,589,933} - 1$ being the largest as at October 2019 [Laroche, 2020]. Further evaluation of the $M_\infty$ asymptote will require a logarithmic transformation of the above formula and more appropriate software.